\documentclass[preprint,aps]{revtex4}
\usepackage{amsmath,amssymb}
\usepackage{epsfig}
\usepackage{graphicx}
\usepackage{dcolumn}
\usepackage{bm}
\usepackage{color}
\usepackage[normalem]{ulem}

\begin{document}

\title{Subconductance states in a semimicroscopic model for a tetrameric pore}

\author{L. Ram\'{\i}rez--Piscina$^1$  and J.M. Sancho$^2$}

\affiliation{$^1$Departament de F\'isica Aplicada, EPSEB, Universitat Polit\'ecnica de Catalunya,
Avinguda Doctor Mara\~n\'on, 44. E-08028 Barcelona. Spain }
\affiliation{$^2$Universitat de Barcelona, Departament de Física de la Mat\`eria Condensada,\\
Mart\'i i Franqu\'es, 1. E-08028 Barcelona. Spain}

\date{\today}

\begin{abstract}
A physical model for a structured tetrameric pore is studied. The pore is modeled as a device composed of four subunits, each one exhibiting two possible states (open and closed). The pore is located within a membrane that separates two reservoirs with ionic solutions. All variables of the model follow physical dynamical equations accounting for the internal structure of the pore, derived from a single energy functional and supplemented with thermal noises. An extensive study of the resulting ionic intensity is performed for different values of the control parameters, mainly membrane potential and reservoir ion concentrations. 
Two possible physical devices are studied: voltage-gated (including a voltage sensor in each subunit) and non-voltage-gated pores. 
The ionic flux through the pore exhibits several distinct dynamical configurations, in particular subconductance states, which indicate very different dynamical internal states of the subunits. Such subconductance states become much easier to observe in sensorless pores.
These results are compared with available experimental data on tetrameric K channels and analytical predictions. 
\end{abstract}


\maketitle

\section{Introduction}

Ion channels are complex biological structures that play a crucial role in controlling ionic transmembrane flow in different cell types\cite{Hille}. These channels are biophysical devices that can permit the flow of specific ions, and whose open and close dynamics can be controlled by membrane voltage, ionic concentrations, and other factors, exhibiting a rich variety of dynamical behaviors. They are involved in many physiological processes, and alterations in their dynamics are associated with a host of physiological disorders, known as channelopathies.\cite{ashcroft2006molecule}

Among the many types of ion channels, voltage-gated potassium (Kv) channels have been extensively studied due to their importance in shaping neuronal excitability, cardiac action potential, etc.\cite{Hammond}
They are associated with diseases such as
type 1 episodic ataxia,\cite{RAJAKULENDRAN2007258}, neuromyotonia\cite{brewer2020structures} or long QT syndrome,\cite{moss2005long} among others.
Kv channels have a tetrameric structure consisting of four modular subunits that control the opening and closing of the pore domain.

Voltage-gated channels are particularly sensitive to changes in membrane potential, which determine not only the magnitude of the ionic flux but also the channel opening and closing dynamics.
This sensitivity together with its tetrameric structure leads to different conductance behaviors, including subconductance states, as the channel can exist in different conformational states depending on the applied voltage \cite{chapman1997activation,bezanilla2005origin,bezanilla2018gating}. In many situations, these intermediate states are rare and difficult to directly observe, but, even in these cases, their presence is indirectly manifest through the activation dynamics, which for instance takes a sigmoidal shape.

The voltage sensor domain in generic Kv channels contains charges or dipoles that are responsible for detecting changes in membrane potential and initiating conformational changes in the pore domain, thus triggering gating events of the channel \cite{catacuzzeno202270}. Nevertheless, there are experiments on other synthesized channels which have no active sensor domain but still they respond to the membrane potential. Comparing both cases, the respective roles of the membrane potential with respect to the gating dynamics are different \cite{syeda2012tetrameric,syeda2016sensorless}.

\begin{figure}[ht]
\includegraphics[width=0.74\columnwidth,clip]{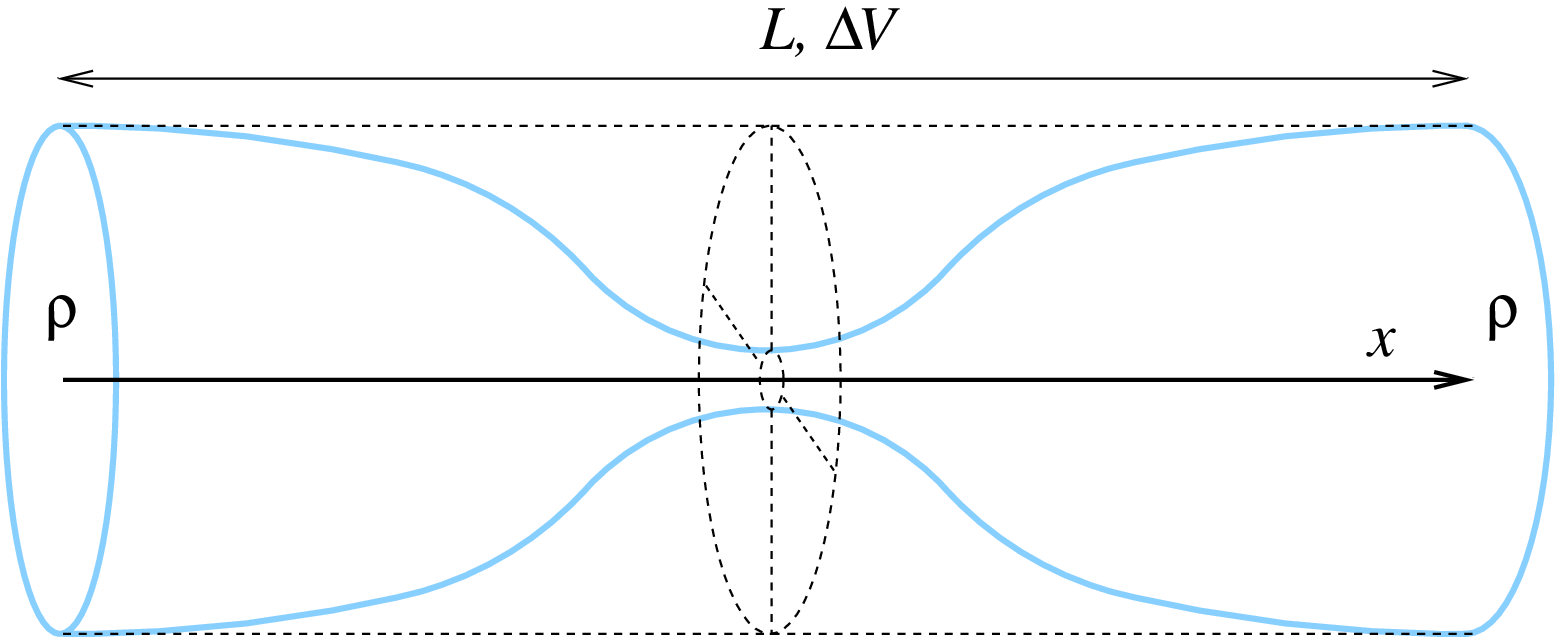}
\includegraphics[width=0.25\columnwidth,clip]{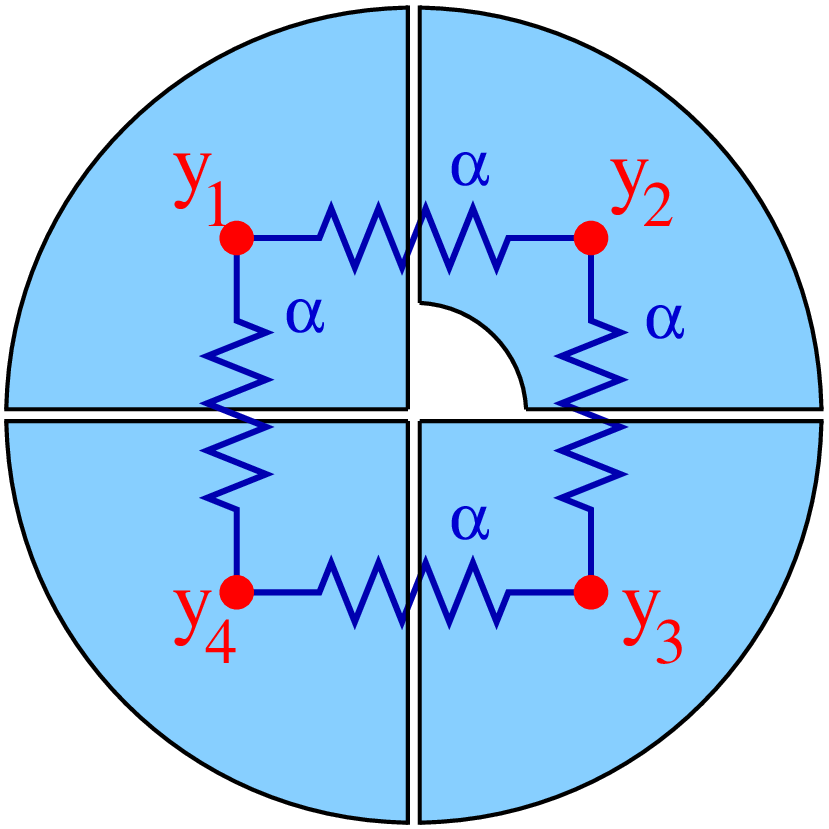}
\caption{Representation of the model tetrameric pore. Left: pore with varying section. Ions move along the $x$ direction. The constriction is modulated by means of the state of the four pore domains. Right: schematics of the four pore domains that reduce the available section for ions. Variables $y_i \in (0,1)$, ($i=1\dots 4$) represent the state of the pore domains. Here three pore domains are in the closed state and one domain ($y_2$) is open. Also the coupling parameter $\alpha$ between neighboring pore domains is represented.}
\label{tretramero}
\end{figure}

The different dynamical open or closed conformational states can be inferred empirically from the ionic flux across the membrane. The analysis of the empirical data concerning these currents can be used to know the different dynamical configurations that the channel can present and, accordingly, can stimulate theoretical models for these devices. In particular, the consideration of results for both sensorless and unmodified channels could help to refine the modeling of the different channel elements. 

In this paper, we focus on a pore model for a voltage-gated Kv channel that has four voltage sensor modules, one in each subunit, which interact with the membrane potential. We have considered an individual dynamics for each of the the four subunits (see for instance Ref. \cite{westhoff2019ks}), but with a coupling between them.\cite{blunck2008fluorescence}
We also consider the deprivation of sensor modules on the same model. We will see how the gating of the channel without sensors is still affected by the value of the membrane potential, an effect observed in some experiments\cite{santos2008molecular} but not explicitly introduced into the model. In this regard previous experimental results indicate that ions can have a direct effect on the channel gating by means of a variety of mechanisms \cite{swenson1981k,holmgren2003influence,elinder2003metal}, some of which have been identified in numerical simulations \cite{aguilella2006blocking,ramirez2}.
In the present modeling, we will see that the presence of ions permits the membrane potential to act on the gating dynamics, even in sensorless channels, by favoring the open states, and that this effect is enhanced by the concentration values.

In Fig. \ref{tretramero} is shown the pore model of length $L$, connecting two reservoirs with equal ionic densities $\rho$, and a section of the four subunits. The state of each subunit is represented by the gate variables $y_j$.  The main control parameters are the membrane potential $\Delta V$ and
the ionic density $\rho$. Dynamics of both ions and gate variables are Brownian, obeying Langevin equations constructed from a common energy functional and with thermal noise verifying the fluctuation-dissipation relationship\cite{ramirez1,ramirez2,ramirez2018periodic}.
By simulation of the model, we can obtain dynamical results such as the ionic flux intensity $I(t)$, the state of the subunits $y_j(t)$, etc. 
In particular, it is convenient to define a conformational order parameter $Y(t)$,
\begin{equation}
 Y = \sum_{j=1}^4 y_j,
\end{equation}
which provides information about the different dynamical configurations of the tetrameric pore, with $Y \in (0,4)$ indicating the number of open subunits.  
From these outputs we compute other quantities such as the frequency of each configuration and their relative probabilities, the relevant time scales, the energy associated with each channel element, etc. 
The details of the model and the numerical methods will be specified below.

\begin{figure}[ht]
\includegraphics[width=\columnwidth,clip]{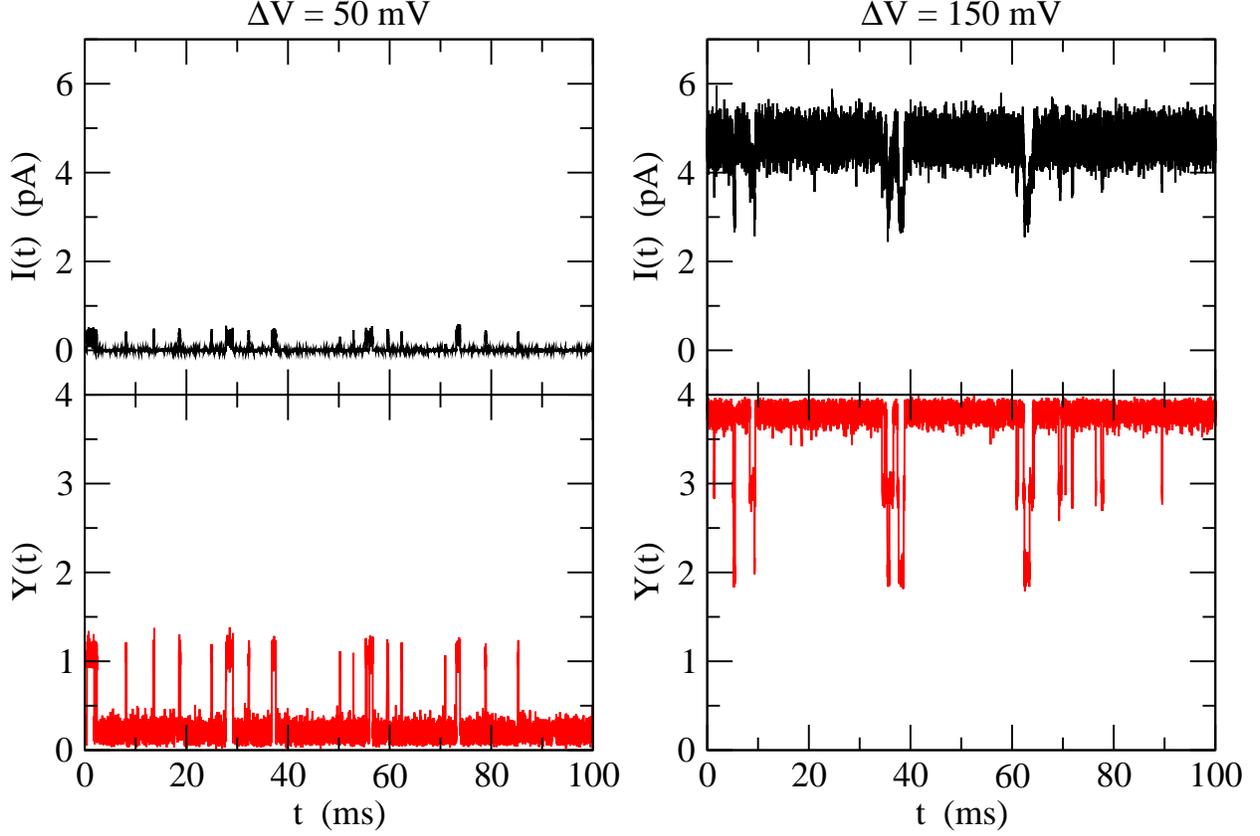}
\caption{Time evolution of a channel with voltage sensor modules (sensor charge $Q=1$e). Top: Ionic flux intensity $I(t)$ filtered with a window of  10 $\mu s$. Bottom: Order parameter $Y(t)$ representing the conformational state of the channel. 
Ion density is $\rho=1$ nm$^{-1}$.  Membrane potential values are: $\Delta V = 50$ mV (left column), and $150$ mV (right column).
}
\label{i-sumy-q1}
\end{figure}

\begin{figure}[ht]
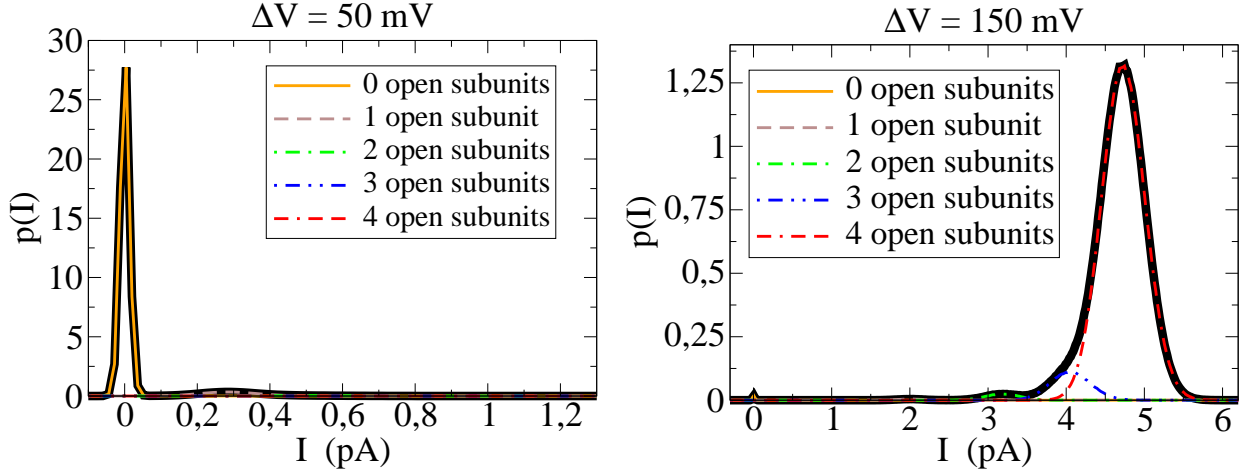

\includegraphics[width=0.48\columnwidth,clip]{I-y-distr-q1-phi50-new-gr.eps}
\hspace{0.5mm}
\includegraphics[width=0.50\columnwidth,clip]{I-y-distr-q1-phi150-new-gr.eps}
\caption{Probability density for filtered intensity values during a long evolution of a channel with voltage sensor modules (sensor charge $Q=1$e), for the same cases as in Fig.~\ref{i-sumy-q1}. Thick black line: for the complete time series; thin colored (gray) lines: for each internal channel state, as derived by the value of the order parameter $Y$.
Membrane potential values are: $\Delta V = 50$ mV (left), and $150$ mV (right).
}
\label{distr-q1-50-150}
\end{figure}

As an illustration of the results obtained from this modeling, we present here some numerical results of a representative case for which pores with and without voltage sensors are compared. 
For the case with voltage sensors, we show in Fig. \ref{i-sumy-q1}, for two different values of membrane potential, numerical results of the output fluctuating current  $I(t)$ (top), and the conformational order parameter $Y(t)$ (bottom), whose value roughly represents the number of open subunits in the channel. In Fig. \ref{distr-q1-50-150} we plot for the same cases the probability density distribution of filtered intensity values, both for the complete time series and for each of the five different dynamical configurations of the tetramer.
The two different membrane potentials that are plotted in both figures are $50$ mV (left) and $150$ mV (right). We see clearly that, most of the time, for the low-voltage case the channel is in a closed estate, and for the larger voltage the channel is completely open. Subconductance states are rare and short-lived, and the channel seems to behave as a single unit with two main conformational states.

\begin{figure}[ht]
\includegraphics[width=\columnwidth,clip]{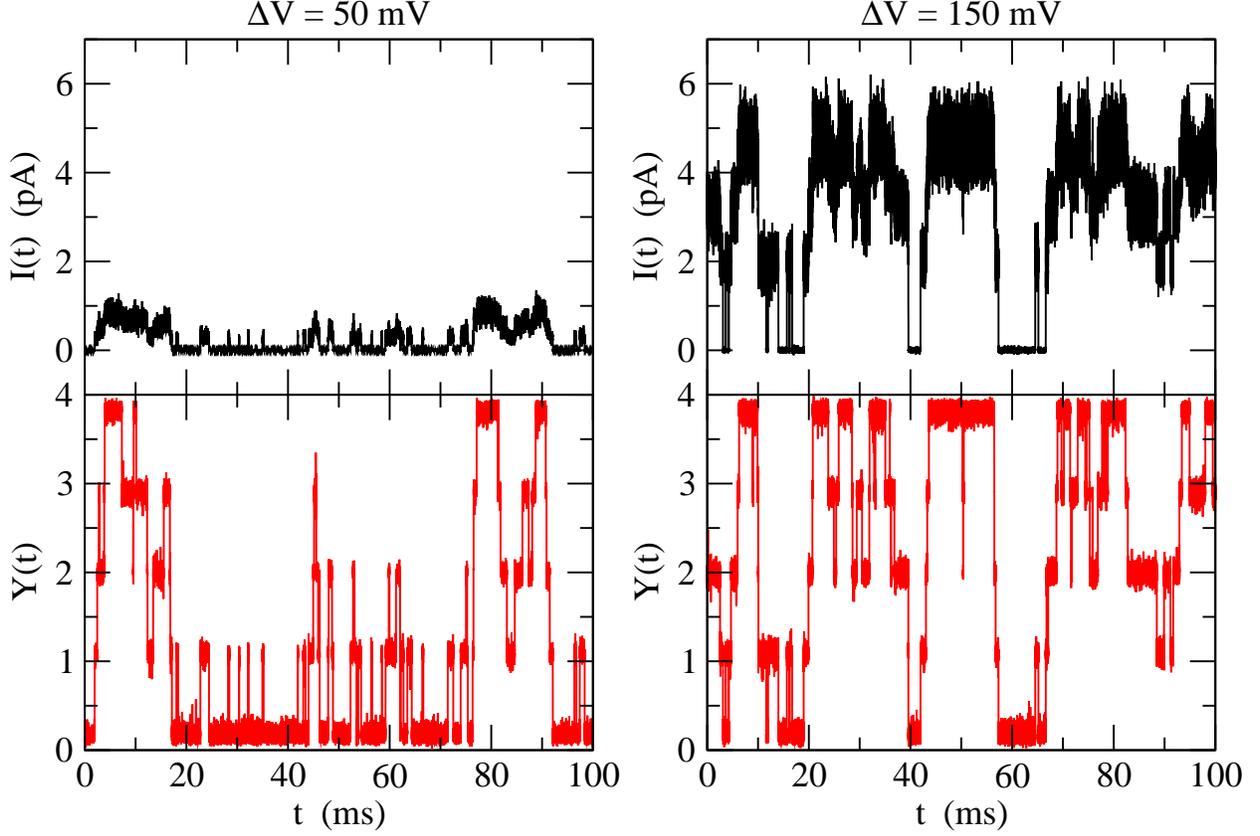}
\caption{Time evolution of a channel without voltage sensors ($Q=0$), with the same notation and parameter values as in Fig.~\ref{i-sumy-q1}.
}
\label{i-sumy-q0}
\end{figure}

\begin{figure}[ht]
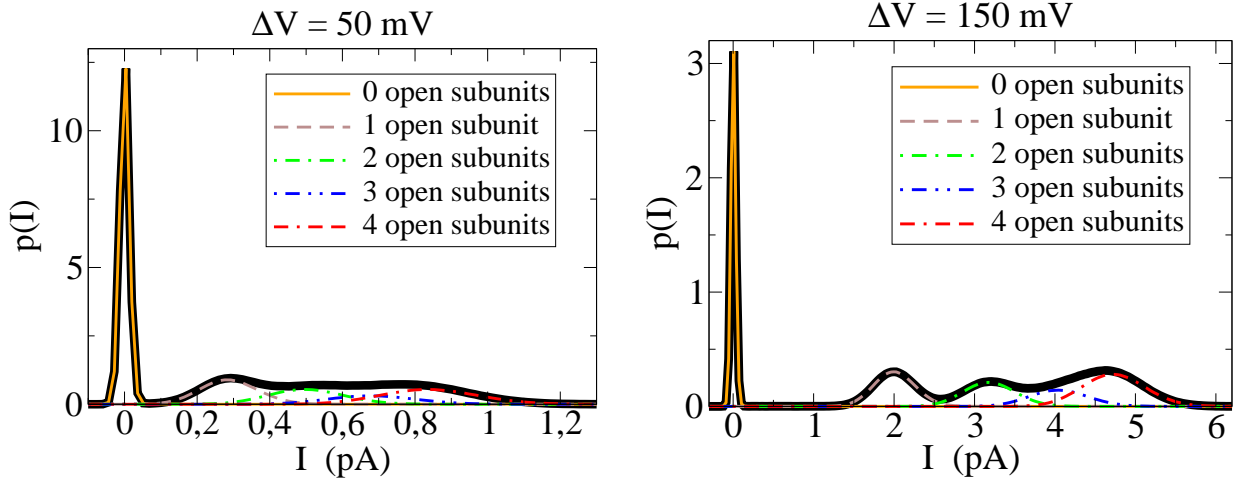

\includegraphics[width=0.48\columnwidth,clip]{I-y-distr-phi50-new-gr.eps}
\hspace{3mm}
\includegraphics[width=0.48\columnwidth,clip]{I-y-distr-phi150-new-gr.eps}
\caption{Probability density distribution of filtered intensity values during a long evolution of a channel without voltage sensors ($Q=0$), with the same notation and parameter values as in Fig.~\ref{distr-q1-50-150}.
}
\label{distr-q0}
\end{figure}

These results can be compared with the case of the sensorless channel as shown in Figs.~\ref{i-sumy-q0} and \ref{distr-q0}. We see that, even in this sensorless case, the increase in membrane potential does have a clear effect on the observed channel states. Still, we can appreciate very important differences with the previous case. The cooperativity in the pore opening, in the sense of a concerted activation of all four channel subunits, is not so strong. Ion current (Fig.~\ref{i-sumy-q0} top) shows multiple dynamical states, including subconductance states, lasting times of the order of milliseconds to tens of millisecond. These states are also observed in the representation of the conformational order parameter, in Fig.~\ref{i-sumy-q0} bottom. The probability density distribution of intensities (Fig.~\ref{distr-q0}) shows different maxima, indicating the presence of several distinct states.  The same effect is observed in experiments with sensorless pore modules\cite{santos2008molecular}.

The outline of this paper is the following. In the next Section II, the physical model of the tetramer is presented, with further details completed in Appendix \ref{app-potential}.
Section III deals with exhaustive data from numerical simulations for sensorless pores, from which a clear phenomenology of their different conformational states is obtained. In parallel, we will compare these results with those of an equivalent tetramer with voltage sensors in order to highlight the differences. The numerical results will be compared with the multiple dynamical behaviors observed in experiments on KbLm molecular channels \cite{chapman1997activation,chapman2005k,santos2008molecular}. Moreover, with the analysis of the output data different dynamical quantities are obtained, with a particular emphasis on the different probability distributions of each conformational state.
Finally, the work ends with a summary of the main results and perspectives.

\section{Methods: The physical model of a tetrameric pore}
\label{methods}

We follow the simplified modeling for a single pore used in Refs.~\cite{ramirez1,ramirez2,ramirez2018periodic}, in which only a reduced set of relevant degrees of freedom is explicitly considered, and the rest of the channel complexity is assumed to be composed by rapid variables whose effect can be reduced to friction and fluctuating terms. This modeling has been modified to take into account the tetrameric structure of the Kv channel. In particular, we have considered the four distinct subunits and their corresponding degrees of freedom.

We consider a 1-dim channel of length $L$, along which the position of each ion $i$, inside the channel, is denoted by $x_i$, $i=1\dots N$, with $N$ being the number of ions. The degrees of freedom of the four subunits are represented by the gate bistable variables $y_j\in (1,0)$  ($j=1\dots 4$, denoting the four subunits), with states $y_j \simeq 0$ and $y_j \simeq 1$ corresponding to the closed and open subunit states respectively. The energy landscape seen by the ions depends both on the membrane potential $\Delta V$ and on the states of the four subunits, $\{y_j\}$, but we neglect direct interactions between ions. 

The dynamical description is based on the use of different energy potentials for each relevant part or mechanism as follows.

\begin{itemize}
\item $V_E(x_i,\Delta V)$ is the potential energy associated with the action of the membrane potential on one ion of elementary charge $e$.

\item $V_I(x,\{y_j\})$ is the interaction between ions and gates. It appears as a barrier potential for the ions, whose height depends on the subunit state.

\item $V_g(y)$ is the bistable potential energy of the gate degree of freedom, with minima at each subunit state (open and closed).

\item $V_s(y,\Delta V)$ is the energy corresponding to the interaction of the voltage sensor module of any subunit with the membrane potential.

\item $V_c(\{y_j\})$ is the coupling energy between subunits. It is simplified by considering only linear interactions between neighboring subunits in a square geometry, as represented in Fig. \ref{tretramero}.

\end{itemize}

Explicit details and parameters of these energy terms are given in Appendix \ref{app-potential}. With all these contributions an energy functional depending on all the dynamical variables is constructed as

\begin{equation}
 U(\{x_i\}, \{y_j\}, \Delta V) = 
 \sum_{i=1}^N V_E(x_i, \Delta V)
 + \sum_{j=1}^4\left[
 V_g(y_j)+ V_s(y_j,\Delta V)
 +  V_c(\{y_j\}) 
 + \sum_{i=1}^N V_I(y_j,x_i) \right]
 \label{pot}
\end{equation}

The stochastic dynamics of ion positions $x_i(t)$  and gate states  $y_j(t)$  is given by the Langevin equations that are obtained from this energy functional \cite{ramirez1} as:
\begin{eqnarray}
\gamma_x {\dot x_i} &=& - \partial_{x_i} U(\{x_i\}, \{y_j\}, \Delta V) + \xi_i(t),\;i=1\dots N
\label{eqx}\\
 \gamma_{y}{\dot y_j} &=&  - \partial_{y_j} U(\{x_i\}, \{y_j\}, \Delta V) +  \xi_{y_j}(t), \;j=1\dots 4
 \label{eqY}
\end{eqnarray}
where the thermal noises fulfill the fluctuation-dissipation  relation, 
\begin{equation}
\langle \xi_a (t) \xi_b(t') \rangle = 2 \gamma_a \,k_B T\, \delta_{a,b}\, \delta (t- t'),
\end{equation}
and $\gamma_a$ are friction coefficients. 

These equations were numerically integrated by a Heun algorithm with a temporal step $\Delta t =5\times 10^{-5}$ $\mu s$.
For each set of parameters, a long run was performed with a simulation time of 10 s, typically taking around 200 h of CPU time on a single core of an Intel-i9 processor at 3.30 GHz.
The ionic concentrations at both sides of the membrane were implemented as boundary conditions at both ends of the channel for the Langevin dynamics of ions \cite{ramirezboundary}.
From the simulation data we recorded the time evolution of ionic flux intensity $I(t)$ and the state of each gate variable $y_i(t)$.
Additional information could be extracted from the evolution of the different variables, such as the probability of each configuration and their corresponding time scales, the interaction energy involved in each configuration, etc.

The values of the employed parameters were: $k_BT=25$ meV, length and main section of the channel $L=4$ nm and $S_0=4$ nm$^2$, ion friction $\gamma_x=0.5$ meV$\cdot\mu$s/nm$^2$, gate friction $\gamma_y=100$ meV$\cdot\mu$s, energy difference between closed and open states of each subunit $D=0.5$  $k_BT$ (slightly favoring the closed state), with a barrier between both subunit states given by $V_0=4$ $k_BT$, reference potential $\Delta V_{ref}=100$ mV, and a coupling energy between contiguous subunits $\alpha= 3$ $k_BT$.

In simulations, we have employed the same ionic concentration value at both boundaries of the channel in order to mimic the conditions of the experiments in Ref.~\cite{santos2008molecular}, to which we have performed the main comparisons. In particular,
we have typically employed a 1-dim density $\rho=1$ nm$^{-1}$ (ions per unit length along the channel, corresponding to a bulk (3-dim) concentration $\rho_\text{bulk}=\rho/S_0=0.415$ M) at both boundaries, but other values have also been used to analyze the effect of ionic concentration. 
 
In each simulation run, the value of the membrane potential has been maintained fixed at the prescribed constant value, in the range $\Delta V = 0\dots 300$ mV. 
The intensity current $I(t)$ is monitored by counting the number of ions crossing the channel during an interval of time, and filtering the results with a window of 10 $\mu$s.
The intensity distributions for each state were obtained by computing their probability density $p(I)$. It is defined such as $p(I)\Delta I$ is the fraction of the total time in which the channel is in the prescribed state and the filtered $I(t)$ takes values in the interval $(I, I+\Delta I)$, for a small $\Delta I$. Its explicit computation was carried out in very long simulations by considering intensity intervals of size $\Delta I = 1.602\times 10^{-2}$ pA, and computing the number $n_i$ of times (from a total of $N$ measurements) that the filtered intensity took values in each of these intervals and the channel was in the corresponding state. By computing the distribution for each interval as $p(I_i)=\frac{n_i}{N\Delta I}$ we obtained a normalized distribution that roughly does not depend on the chosen $\Delta I$.
Specifically the area under the curve of any state (colored thin lines in Figs.~\ref{distr-q1-50-150},\ref{distr-q0}) is equal to the probability of this state, and the total area of the complete distribution (thick black line in the same figures) is equal to unity.

We have then explored the role of both $\rho$ and $\Delta V$ parameters in a non-voltage-gated tetramer (sensor charge $Q=0$). Comparisons with a voltage-gated channel ($Q=1$e) have also been included.
In the initial stages of the work, we also explored the role of other parameters, defining the model but which cannot be easily changed in experiments, such as $D$, $\alpha$, $\beta$, $V_0$, $\gamma_x$ and $\gamma_y$, in an ample range.
Results (not shown here for the sake of clarity) did not change qualitatively, and the observed changes in these parameters only affected results in expected aspects like temporal scales, relative weights of the different states, etc., not affecting the main conclusions of this work and thus confirming the robustness of the model. Such explorations should also permit one to tune the parameters for a better quantitative correspondence with specific experiments.
Other tetrameric channels could also be considered, within this modeling, with the appropriate modifications and, possibly, different values of these parameters.

\section{Results}
 
 The numerical output of the model variables $x_i(t),y_j(t)$ is computed to get the physical variables: ionic flux intensity $I(t)$ and order parameter $Y(t)$. With this information, different and useful physical results are obtained which are described below.  These results are ordered as dependent on the two external control parameters: membrane potential $\Delta V$ and ionic density in the reservoirs $\rho$.
 
 \subsection{Role of the  membrane potential $\Delta V$}
\label{rolepotential}

{\bf Pore dynamical configurations}. In the previous Figs.~\ref{i-sumy-q1}-\ref{distr-q0} two different membrane potentials were employed for both sensorless and voltage-gated pores. As already discussed above, the channel with sensors (Figs.~\ref{i-sumy-q1},\ref{distr-q1-50-150}) essentially shows fluctuations around two different values of the intensity, corresponding to the open and closed channel states, reached depending on the membrane potential.
However, the sensorless channel (Figs.~\ref{i-sumy-q0},\ref{distr-q0}) presents, for each membrane potential, multiple observable steady states, with $I(t)$ fluctuating around different values. The intensity values are correlated with the different values taken by the order parameter $Y(t)$, observing a distribution of relative maxima of the probability density $p(I)$, which corresponds to the five possible dynamical configurational states of the pore.  It is worth commenting that we observe in $p(I)$ four maxima instead of five, because stochastic fluctuations make the distributions of intensities wider, and not all the states are isolated enough to be resolved.  A similar behavior, but in an experimental setup, is observed in Fig. 6 of Ref. \cite{chapman1997activation}.  In the experiments of Ref. \cite{santos2008molecular} these five states are not clearly seen either (see for instance Fig. 9 of that reference).

\begin{figure}[ht]
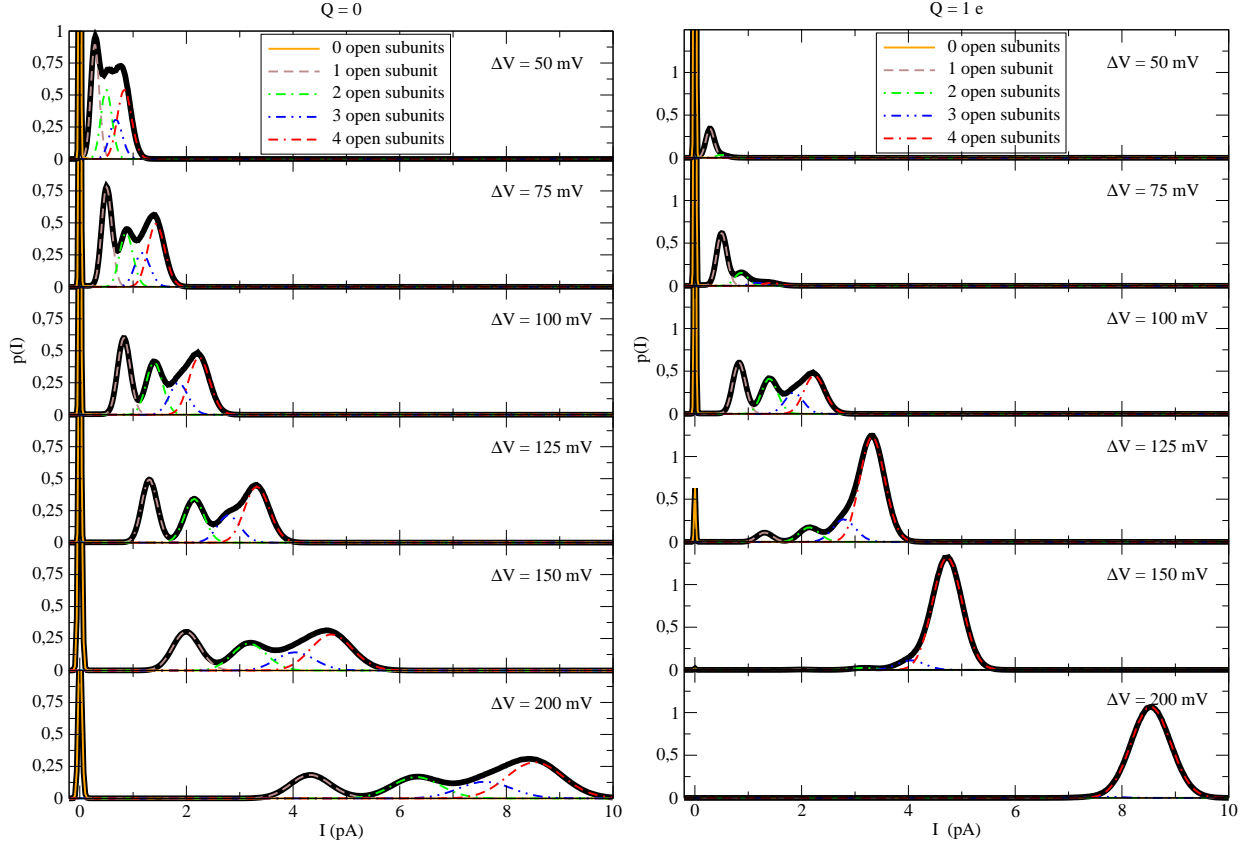

\includegraphics[width=0.497\columnwidth,clip]{y0.5alpha3-distrib-I-y-new4-gr.eps}
\includegraphics[width=0.49\columnwidth,clip]{q1-distrib-I-y-new4-gr.eps}
\caption{Probability density distributions of filtered intensity values. Left: channel without sensors ($Q=0$); right: channel with sensors ($Q = 1$e, $\Delta V_{ref}=100$ mV). $\rho = 1$ nm$^{-1}$ in both cases. Thick black line: for the complete time series; thin colored (gray) lines: for the times corresponding to each internal channel state, as derived by the value of $ Y=\sum y_j$. In order to use the same scale for all voltages, the large first peak near $I=0$ has been cut in almost all cases.}
\label{q0-q1-distrib-I-y}
\end{figure}

\begin{figure}[htp]
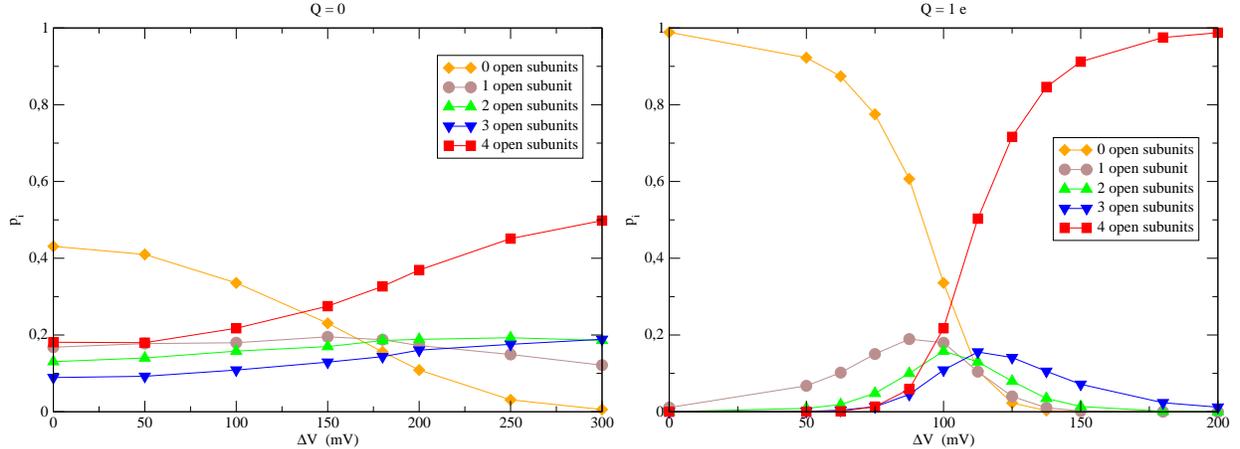

\includegraphics[width=0.49\columnwidth,clip]{p-y0.5-a3-new3.eps}
\includegraphics[width=0.49\columnwidth,clip]{p-y0.5-a3-q1.eps}
\caption{Probability for each internal state (computed as the time fraction in each state) vs membrane potential 
$\Delta V$. Left: sensorless channel ($Q=0$); Right: channel 
with sensors ($Q=1$e). $\rho=1$ nm$^{-1}$ in both cases. Different symbols and colors (gray levels) indicate the channel state (see insert).}
\label{p-y0.5a3}
\end{figure}

We analyze in more detail the dependence on the membrane potential in Fig.~\ref{q0-q1-distrib-I-y}, where we employ six different values of $\Delta V$, and the two types of channels can be compared. The first obvious effect is the (trivial) displacement of the intensity distributions towards higher values as the membrane potential is increased. Clearly, the intensity crossing the channel in a given configurational state should be higher as the potential increases. More interesting is the presence of subconductance states, which appear to be more persistent in sensorless channels ($Q=0$) for all potentials tested. For this case, we see that the states with a larger number of open subunits tend to take a greater statistical weight when increasing potential, while those with more closed subunits tend to lose statistical weight. Similar distributions are found in the experiments with sensorless channels of Ref.~\cite{santos2008molecular}, where a dependence on the applied voltage is also observed. A more quantitative analysis is performed if Fig.~\ref{p-y0.5a3}-left, where the probability of each conformational state is represented as a function of membrane potential. In this figure we reach even higher voltage values. We see that all states have a non-zero probability at zero potential, with the all-closed state (0 open subunits) being the state with the higher probability. This very state reduces its probability as the voltage increases, reaching values close to zero for voltages near 300 mV. The rest of the states initially increase their probability, although the subconductance states seem to reach a maximum at some intermediate voltage: $\sim 150$ mV for the 1 open subunit state, and $\sim 250$ mV for the 2 open subunits state; the 3 open subunits state seems to have much reduced the slope at the largest tested voltage so it seems reasonable to also expect a maximum for some larger value of the membrane potential. The probability of the all-open state is observed to monotonously increase with voltage, but it is apparently very far from saturation even for the largest tested voltage.

This is in contrast with the voltage-gated channel ($Q=1$e), where the dynamics is dominated by two main states (all closed and all open states), and the subconductive states are more rarely observable. In Fig.~\ref{q0-q1-distrib-I-y}-right there still can be seen local maxima of subconductance states for some intermediate voltages.
However, we must bear in mind that in the first three cases of this figure (voltages up to 100 mV) the peak at $I=0$ is very large (it has been cut to see the characteristics of the rest of the distribution), so in fact in these cases the closed state dominates.
For larger voltages, it is the all-open state which dominates and takes most of the statistical weight. This is best seen in Fig.~\ref{p-y0.5a3}-right, where both main states present sigmoidal probabilities, dominating the dynamics, and subconductance states are significantly less probable except for a limited range of intermediate values of the membrane potential. Subconducting states also present maxima at intermediate values, with positions slightly shifted to higher values of voltage as the number of open subunits is larger.

It is remarkable that, in the $Q=0$ case, no explicit interaction of the membrane potential with the gating variables has explicitly been included in the model equations. 
This observed effect is indirect and, by construction of the model, it can only be mediated by the ions.
Indeed, the interaction term $V_I(Y_j,x_i)$ in Eq.~(\ref{pot}) between ions and gates depends on both ion and gate variables, which implies that it appears in the dynamic equations for both types of variables. That is, when gates exert forces over the ion, the ion also acts over the gates, which is a manifestation of the 3rd Newton's law. Since both forces come from the same potential energy, and the involved degrees of freedom evolve with the thermodynamically consistent formulation of Eqs.~(\ref{eqx},\ref{eqY}), with the correct thermal noises, we expect this effect to be real, even in the context of such a crude modeling of a channel.

\begin{figure}[ht]
\includegraphics[width=0.8\columnwidth,clip]{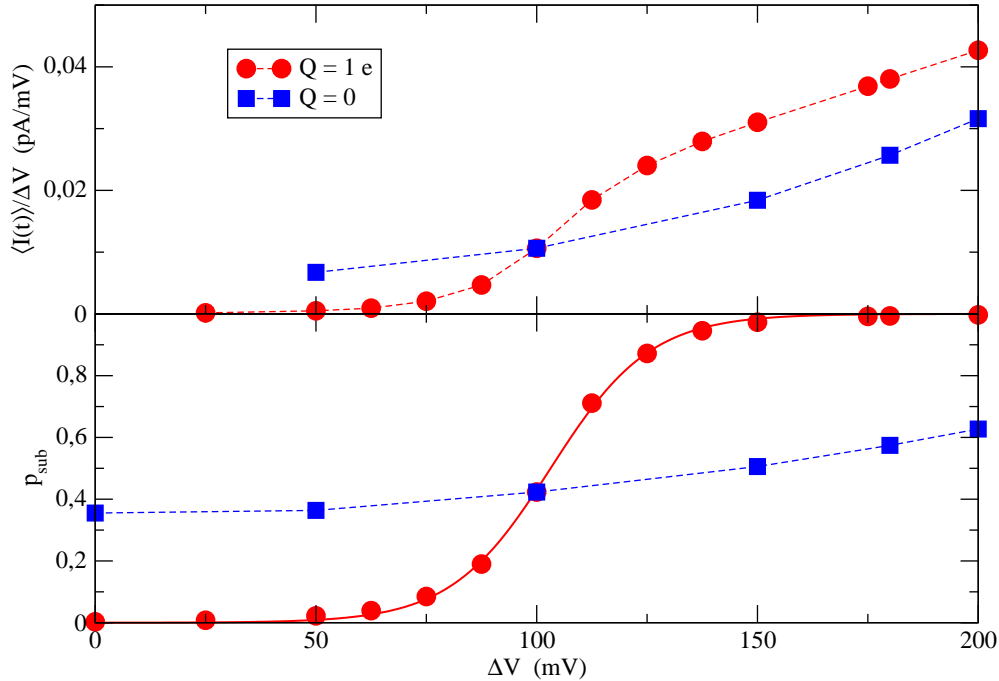}
\caption{Channel conductivity ($\langle I(t)\rangle/\Delta V$, where the mean value corresponds to a temporal average) (top) and open probability of each channel subunit (bottom), as a function of membrane potential. Red circles: channel with sensors ($Q=1$e for each subunit); blue squares: sensorless channel ($Q=0$); red solid line: fit of the open probability for subunits with sensors (see Eq.~\ref{fit-p-q1}). Dashed lines are guides to the eye.
}
\label{conduct-prob1-y0.5a3-q0-q1}
\end{figure}

{\bf Pore conductivity and subunit open probability}. 
In Fig.~\ref{conduct-prob1-y0.5a3-q0-q1}-top we present results for the effective channel conductivity, represented by the quotient $\langle I(t)\rangle/\Delta V$, vs. the applied membrane potential. The mean value is computed as a temporal average along the simulation time, and the results are then associated with the gating dynamics. They can also be related to the permeability of a portion of membrane with a large number of channels. 

In the case of the voltage-gated channel ($Q=1$e) we clearly see two regimes. For low voltages, the conductivity is very low, whereas for large voltages the conductivity is larger, with a nearly linear dependence. There is a crossover between both regimes at intermediate voltages, where the slope of the $\Delta V$ dependence is larger. This behavior can be understood as a manifestation of the channel bistability, where the two regimes correspond to the closed and open states adopted for low and high voltages respectively. The crossover is placed at the region in which gating occurs. 
On the contrary, 
bistability is not so apparent in the conductivity of the sensorless channel ($Q=0$), where the changes are more gradual and only a smooth increase is observed when increasing the membrane potential.

We can relate the effective conductivity of the channel with its gating activity. To this end we have computed $p_\text{sub}$, the probability of the open state for any subunit, by evaluating the proportion of the total time in which each subunit remains open in long simulation runs. We can see the results in Fig.~\ref{conduct-prob1-y0.5a3-q0-q1} (bottom). We can appreciate some similarities between the dependences of conductivity and open probability on the voltage in each case.
In particular, for the sensorless case, we see a weak but non-negligible dependence on voltage, thus confirming that the membrane potential affects the gating mediated by the ion current.

For the case of a voltage-gated pore, the open probability for the channel subunits presents a sigmoidal shape marking the crossover between both states of the subunit. It has been fitted as \begin{equation}
 p_{sub}(\Delta V) = \frac{1}{2} \left(1 + \tanh\left[\frac{Q_\text{eff}(\Delta V-\Delta V_\text{eff})}{2k_BT}\right]\right).
 \label{fit-p-q1}
\end{equation}
This function corresponds to the Boltzmann probabilities of a bistable gate with an effective sensor charge $Q_\text{eff}$, and has often been used in the analysis of channel gating.
The result of the fit is: 
$Q_\text{eff}=2.21$ e, $\Delta V_\text{eff}=103$ mV. 
The value of $\Delta V_\text{eff}$ compares well with the value used in simulation $\Delta V_\text{ref}=100$ mV (see Eq.~(\ref{gateequ}) in Appendix \ref{app-potential}). 

Regarding the value of the effective sensor charge $Q_\text{eff}$, we note that the value should depend directly on the coupling between subunits (with likely a very small correction due to the presence of ions). 
On the one hand, for the case of completely uncoupled subunits (parameter $\alpha=0$), in which each one is affected by the membrane potential and they are independent from each other, one would expect an effective sensor charge close to $Q=1$e, that is the value used in simulations for each subunit. On the other hand, for the completely opposite case of very large coupling ($\alpha \rightarrow  \infty$), one would expect the four subunits to behave as a single gate with a total charge $4Q=4$e, and $Q_\text{eff}$ would be close to this value. Then a monotonous dependence of $Q_\text{eff}$ on $\alpha$ is expected, and the obtained value has indeed been intermediate between both limits. 
This indicates that the analysis of $p_\text{sub}(\Delta V)$ in experiments and the computation of the corresponding  $Q_\text{eff}$ for subunits could provide valuable information about the coupling between subunits in real tetrameric channels.

\begin{figure}[ht]

\includegraphics[width=0.8\columnwidth,clip]{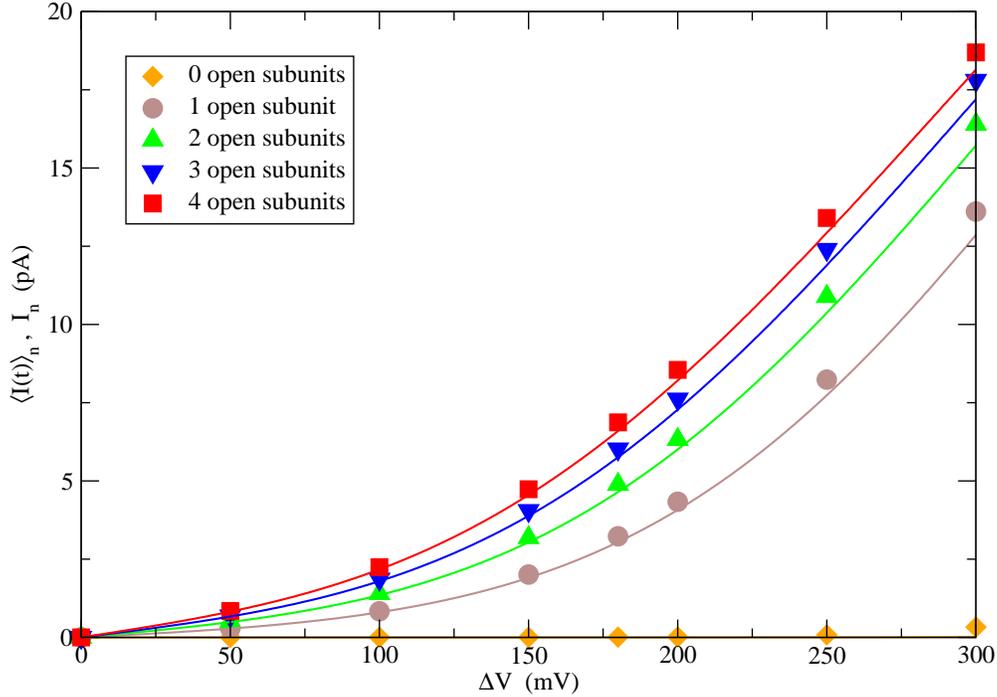}
\caption{Mean intensities for each internal state of the channel vs membrane potential. Symbols: $\langle I(t)\rangle_n$ from simulation results, averaged during each internal state of the channel; lines: theoretical results for $I_n$ (Eq.~\ref{theo-I-V}). The density in all cases is $\rho=1$ nm$^{-1}$. Different symbols and colors (gray levels) indicate the channel internal state $n$ during the averaging (see insert).}
\label{int-y-y0.5a3}
\end{figure}

{\bf Ionic intensity in each configurational state}. 
We have seen how gating plays a fundamental role by modulating ion intensity. To subtract the effects of gating, we obtained the mean intensities for each internal state as a function of the membrane potential. The results are presented in Fig.~\ref{int-y-y0.5a3}, and correspond essentially to the position of the peaks in Fig.~\ref{q0-q1-distrib-I-y}. As expected, intensities increase with potential in all cases. Moreover, the slopes also increase, indicating an increase in the conductivity at very high voltages. Note, however, that for such large values of the membrane potential
other effects not included in the model could possibly enter into play.\cite{nelson2003modeling}

These simulation results have been compared to the classical analytical expressions for the flow of Brownian particles through a pore. Considering $V_n(x)$ to be a (static) potential  for each ion, moving with a Langevin dynamics, with $n$ open subunits, the resulting intensity is (see Eq.~(\ref{theory-I-V}) and derivation in Appendix \ref{app-intensities})
\begin{equation}
 I_n = I_0 \frac{e^{V_n(0)/k_BT}- e^{V_n(L)/k_BT}}{ \int_0^1 e^{V_n(zL)/k_BT} dz},
 \label{theo-I-V}
\end{equation}
where, for the parameters of our simulations, the value of the constant $I_0 = 2.0025$ pA. To plot this prediction, we have considered for the potential $V_n(x)=V_E+V_I$, that is the terms in Eq.~(\ref{pot}) affecting a single ion, and furthermore considering constant ({\it i.e.} no fluctuating) values for the gate variables $y_j$ depending on the channel state. We find in Fig.~\ref{int-y-y0.5a3} a good agreement, which means that this static approximation, together with the analytical tools from the stochastic processes theory, could be of use in the analysis of this kind of problems.

\subsection{The role of ionic concentration in sensorless channels ($Q=0$)}

In Subsec.~\ref{rolepotential} we have shown that membrane potential has a clear effect on the gating of a sensorless channel. As already discussed, in our modeling this effect could only be mediated by the ions.  In this section, we study the dependence of gating on the value of the ionic concentration in these channels.

\begin{figure}[ht]
\includegraphics[width=0.8\columnwidth,clip]{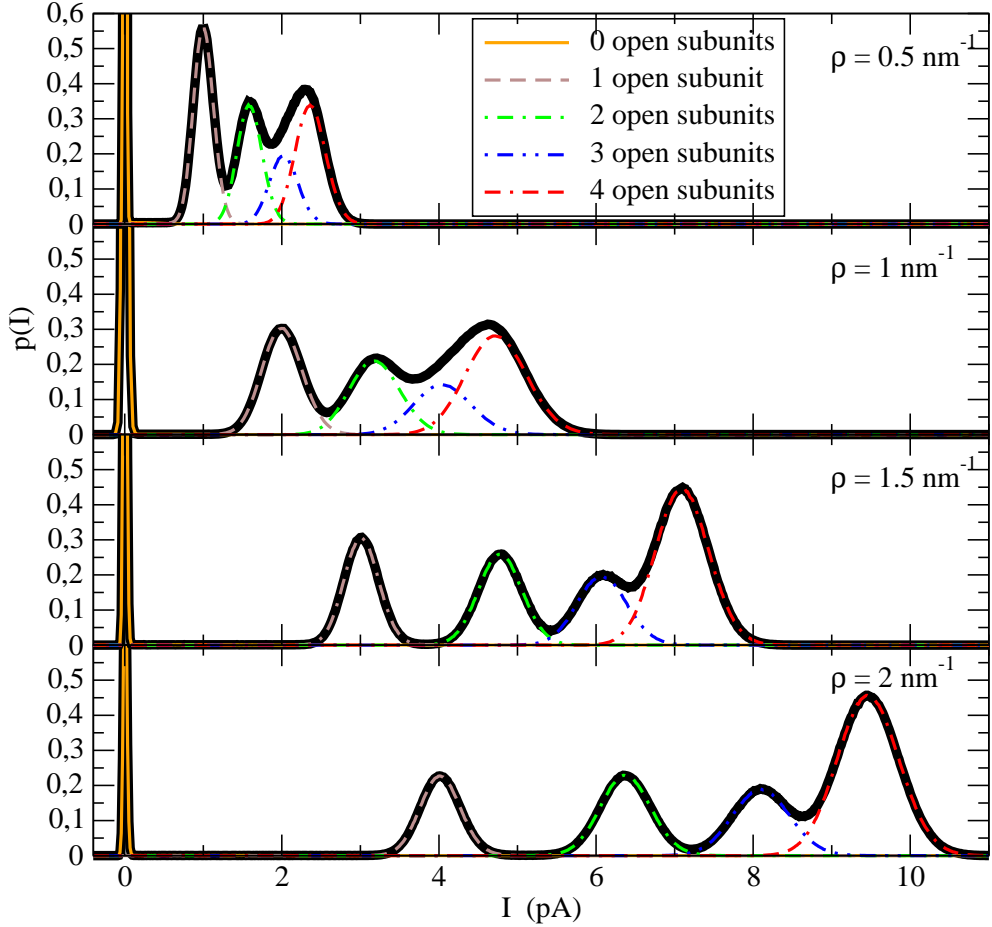}
\caption{Probability density distributions of filtered intensity values for  $\Delta V = 150$ mV and different concentrations. Thick black line: for the complete time series; thin colored (gray) lines: for the times corresponding to each internal channel state, as derived by the value of $ Y=\sum y_j$.}
\label{y05phi150alpha3-distrib-I-y-rho}
\end{figure}

{\bf The pore configurations}.  In Fig. \ref{y05phi150alpha3-distrib-I-y-rho} we observe different intensity distributions corresponding to four increasing concentrations, maintaining a constant value of the membrane potential $\Delta V = 150$ mV.
First, we see that the position of the peak for each internal channel state is proportional to the concentration value. This is a direct consequence of the absence of ion-ion interactions in the model. More interesting is the fact that the relative distribution of weights of these peaks changes, which is a manifestation of the effect of the concentration on the gating. In particular, it is manifested that the increase in concentration enhances the probabilities of the more open configuration while hindering the more closed ones. We see also that for lower concentrations, by observing intensity values, the distributions of the different states are more difficult to separate and distinguish, and that one of the subconductance states is completely hidden. However,
for larger concentrations, the five configurations are more separated and can be clearly seen.

\begin{figure}[ht]
\includegraphics[width=0.6\columnwidth,clip]{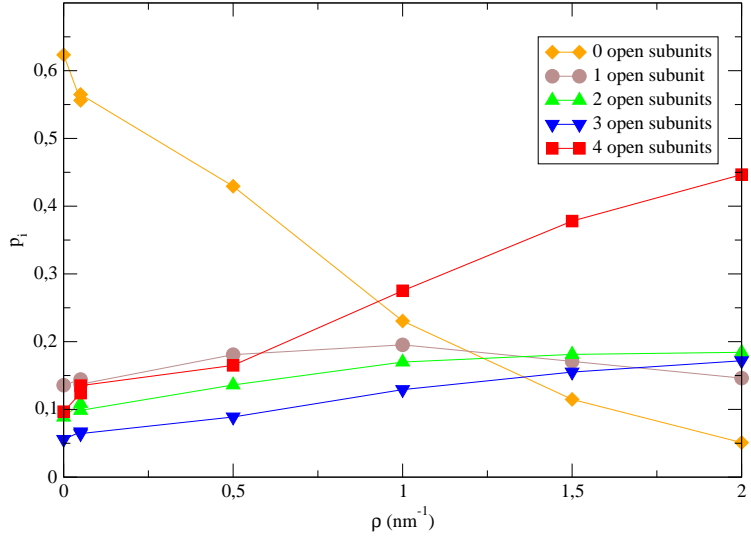}
\caption{Probability for each internal state (computed as the time fraction in each state) vs ion concentration $\rho$. $\Delta V= 150$ mV. Lines are a guide to the eye.}
\label{d05-phi150-alpha3-p-y-rho}
\end{figure}

This trend is quantitatively confirmed with the calculation of the total probability for each internal channel state shown in Fig. \ref{d05-phi150-alpha3-p-y-rho}, where they are plotted versus the ion density $\rho$.
We see here that the probability for the all-open configuration increases monotonously with concentration, while for the all-close one it decreases. The probability of the intermediate states has a much weaker dependence. Furthermore, the intermediate state with 1 open subunit presents a smooth maximum around $\rho = 1$ nm$^{-1}$ and, for the other intermediate states, the change in slope appears also to indicate the likely presence of a maximum at higher values of $\rho$ as the number of open subunits is larger.
In fact, the dependence on density shown in this figure appears to be very similar to the dependence on voltage depicted in Fig.~\ref{p-y0.5a3}-left for the same channel. The conclusion is that the role of concentration on the gating of sensorless channels is parallel to that of the voltage, and this is associated with the fact that the permeant ions, interacting with the barriers, constitute the mechanism by which the membrane potential acts on the gating in this modeling of sensorless channels.

\begin{figure}[ht]
\includegraphics[width=0.6\columnwidth,clip]{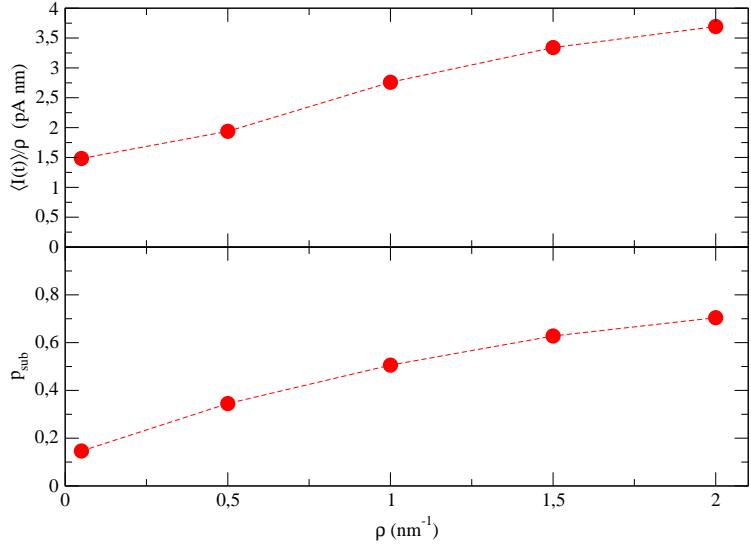}
\caption{$\langle I(t)\rangle/\rho$ (top) and open probability of each channel subunit (bottom) for sensorless channels as a function of $\rho$. $\Delta V = 150$ mV. Lines are a guide to the eye.
}
\label{conduct-prob1-y0.5a3-rho}
\end{figure}

{\bf The intensity}. 
The ratio $\langle I(t)\rangle/\rho$ in shown in Fig.~\ref{conduct-prob1-y0.5a3-rho}-top.
From this figure it can be checked that the resulting mean intensity $\langle I(t)\rangle$ is not simply proportional to the concentration $\rho$ (the ratio $\langle I\rangle/\rho$ is not constant), as it was for each internal state. This is obviously an effect of the gating, due to the increase of probability of the open states with ion concentration. 
This probability, specifically the open probability for each subunit, in shown in Fig.~\ref{conduct-prob1-y0.5a3-rho}-bottom, and it is a monotonously increasing function of $\rho$, showing a dependence very similar to that of $I/\rho$.

A similar effect of ion concentration on gating was already demonstrated with a simpler (two states) channel model in Ref.~\cite{ramirez2}, where it was also shown that the increase of ion concentration favored the open state. It is worth mentioning that in that model a similar formulation, using a single energy functional for all interactions, was also employed.

\section{conclusions}

We have presented a simplified model for a tetrameric ion channel, in which only a reduced set of variables (ion positions and the four degrees of freedom of the pore subunits, acting as gates) have been considered, all evolving according to Brownian dynamics, with interactions defined by a single energy functional and noises obeying the fluctuation-dissipation relationship.
Both voltage-gated and sensorless pores have been considered. In the model for the sensorless pore the four bistable subunits have a weak coupling between them, but no direct interaction with the membrane electrostatic field. Ions interact with both membrane potential and pore subunits. In the case of the voltage-gated pore the same model and parameter values have been employed for the sake of comparison, but with the addition of the interaction between a sensor (one elemental electric charge in each subunit) and the membrane potential.

Simulations of the voltage-gated pore for different values of the membrane potential have shown that the model behaves as a bistable device where, depending on the voltage value, the two main channel states (with all subunits either closed or open) dominate the dynamics. Intermediate (semiconducting) states are sporadic and short-lived, and are only present in a significant way for a limited range of intermediate values of the voltage.
The open probability for each subunit presents the expected Boltzmann dependence on the voltage for a bistable system, with an effective charge slightly larger than twice its real charge. Since this effective charge is expected to depend on the coupling between subunits, a more detailed study could in principle permit one to obtain information on this coupling by analyzing experiments on real tetrameric channels. Furthermore, results for the ion flow achieved during each state (during which the channel variables still fluctuate) agree well with the classical prediction for the flow of Brownian particles along a static potential, probably due to the difference of time scales between ions and gates. This opens the way to study this problem by using a static approximation \cite{ramirezstatic} that could accelerate simulations and permit the consideration of a large number of channels. That could be useful to relate simulation results to physiological conditions.

Results of simulations of the sensorless channel compare qualitatively well with the experiments on genetically modified sensorless channels of Ref.~\cite{santos2008molecular}. In this case, the subconducting states (with an intermediate number of open subunits) have been more probable and definitively observable by monitoring for instance the ionic flow values. Moreover, the channel is affected by the membrane potential, but, instead of appearing as a bistable device like in the voltage-gated case, it responds to it more smoothly. In particular, results have shown that the membrane potential has a clear effect on the distribution of intensity values and the relative frequency of the different conformational channel states, in a way very similar to what was observed in the experiments of Ref.~\cite{santos2008molecular}. That is, the electric field is able to act on the gating without any voltage sensor. In this action the increase of the voltage has the general effect of favoring the open state of the subunits, thus altering the distribution of intermediate states towards the more open states of the channel. As a consequence, the mean conductivity of the channel is enhanced. 

This effect of the membrane potential on the gating of the sensorless pore is attributed to the presence of ions. This has been confirmed by the simulations with different values of ion concentration, where the changes have been analogous to those of variations of the voltage values. Similar results have been obtained in simpler models of gated pores.\cite{ramirez2} The direct action of ions on the gates is an unavoidable consequence of the physically consistent formulation of the model using a single energy functional. This permits the mutual action between ions and gates, in a form of Newton's third law. In general, experiments on real channels have permitted the identification of several distinct mechanisms accounting for a variety of different ion effects of gating.\cite{swenson1981k,holmgren2003influence,elinder2003metal} Our results have shown that, for explaining some specific observed effects, the direct interaction between ions and gates could be sufficient.

The model presented here could be useful for studying the dynamics of subconducting states in other situations. An interesting one corresponds to the experiments of Ref.~\cite{chapman2005k}, in which such states could be controlled by the synthesis of pores formed by subunits with very different activation voltages. Also, in Ref.~\cite{syeda2012tetrameric}, the number of voltage sensors in the tetramer could be controlled by combining a variable number of full-length subunits and sensorless pore modules. Our model would be most appropriate for analyzing these kinds of heterotetramers. More detailed data on similar experiments could also be interesting for quantitative comparisons with our model.

A similar formulation could also be used for modeling other channels, such as the Na or Ca channels. The tetrameric structure of the Na channel, for instance, is not symmetrical, since the four voltage sensor domains forming its $\alpha$ subunit are not equivalent,\cite{ahern2016hitchhiker,goldschen2013multiple} situation that could be addressed within the present approach.
It is also interesting to note that some characteristics of the activation dynamics of many channels depend on the existence of the intermediate states, even if such states are not directly observable, and could be reproduced with an analogous modeling.
For instance, BK channels very rarely transit to subconductance states, but still these states play an important role in the gating events.\cite{magleby2003gating}
In this regard, this model could be useful, with the appropriate modifications, for describing the dynamics of many different types of tetrameric ion channels.
Finally,
the model could also be completed with additional electrophysiological details for studying processes at the scale of each gating event, such as the dynamics of the gating currents,\cite{catacuzzeno2019simulation} in tetrameric pores.

\begin{acknowledgments}
L.R.-P. acknowledges financial support through
Grant PID2022-139215NB-I00, funded by MCIN/AEI/10.13039/501100011033 (Ministerio de Ciencia e Innovación, Spain) and by “ERDF: A way of making Europe” (European Union),
and through
Grant 2021-SGR-00582, funded by Agència de Gestió d’Ajuts Universitaris i de Recerca (Generalitat de Catalunya, Spain).
J.M.S. acknowledges financial support from the Spanish Ministerio de Ciencia, Innovación y Universidades / Agencia Estatal de Investigación / Fondo Europeo de Desarrollo Regional, Unión Europea through grant PID2021-125202NB-I00.
\end{acknowledgments}

\bibliography{channels.bib}

\clearpage

\begin{appendix}
\section{Potential energies}
\label{app-potential}

In this appendix we provide details on the different terms appearing in the formulation of the energy functional defining the model in Sec.~\ref{methods}. Each term is the potential energy corresponding to the interactions associated with each of the physical mechanisms considered in the model.

{\bf -Interaction ion-membrane potential.} 
The energy associated to the interaction between the membrane potential $\Delta V$ and an ion of elementary charge $e$, at position $x$ along the channel with length $L$, is written as
\begin{equation}
 V_E(x_i,\Delta V)=-\Delta V \frac{x_i}{L}.
\end{equation}

{\bf -Interaction between ions and gates}. The interaction between each ion and the pore gates is described by an effective potential, depending on both the ion position $x_i$ and the states of the four subunits.
These states are described by the value of four variables $y_j$  ($j=1,2,3,4$), representing the degrees of freedom of the four subunits forming the tetrameric structure ($y\sim 0$ closed, $y\sim 1$ open).
The potential then includes the entropic effect of varying the section available for the ions along the channel. According to Ref.~\cite{zwanzig1992diffusion}, the effective entropic potential along a channel with a varying section $S(x)$ and opening $S_0$, for a particle at position $x$, is given by $V(x)=-k_BT\ln(S(x)/S_0)$. In our case, we model the channel with a constriction in the middle part, of length $\sigma$, controlled by the parameter $\beta$ which corresponds to the fraction of the section that is reduced at the constriction when the channel is open. The available section for the ion can be further reduced by the gates, depending on the subunit states $\{y_j\}$, when they close.

\begin{figure}[ht]
\includegraphics[width=0.6\columnwidth,clip,angle=0]{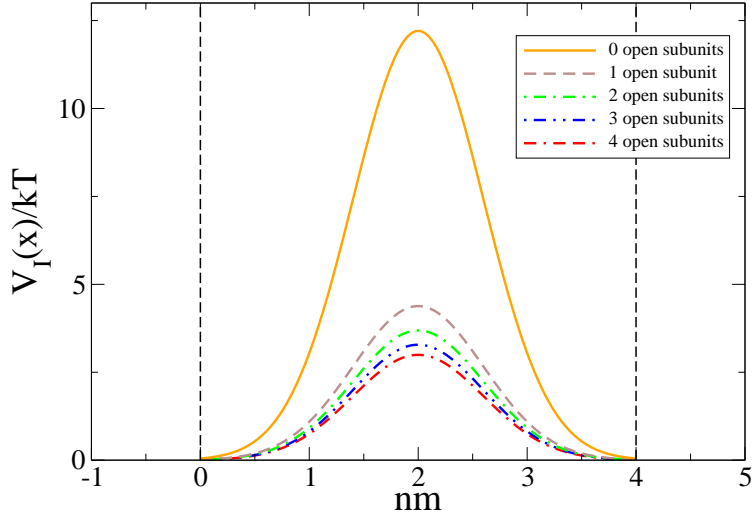}
\caption{Potential barriers corresponding to the five different conformational states of the tetramer (see inset).}
\label{barreras}
\end{figure}

The resulting potential takes the form:
\begin{equation}
 V_I(x,\{y_j\}) 
 =-k_BT e^\frac{-(x-\bar x)^2}{2\sigma^2}\ln[(1-\beta)(1-{\frac{1}{4}}\sum_{j=1}^4f(y_j))].
 \label{VS}
\end{equation}
Here the function 
\begin{equation}
f(y)=\frac{1}{2}(1-\tanh(y-\frac{1}{2})/0.1)
\label{eq:f}
\end{equation}
is used to reduce the sensitivity to thermal fluctuations on the steady states near $y\sim 0,1$. For the present simulations, the constriction is characterized by $\beta=0.95$, a length scale $\sigma = 0.15 L$, and a position $\bar x = L/2$, that is it is at the center of the channel. The resulting potentials for the different conformational states of the channel are shown in Fig.~\ref{barreras}.

{\bf -Energy potential of a gate}.  
The state of the pore is given by the value of the four variables $y_j$.
For each $y_j$ variable, a bistable potential is defined,
\begin{equation}
 {\textstyle V_g(y)=V_0[-a\ln[y(1-y)]-b(y-\frac{1}{2})^2]+Dy},
\end{equation}
where we use the values $a=0.1$, $b=15$. Then the dynamics for the $y_j$ variables is limited to the interval (1,0). There are two minima very close to both limits 
(specifically for $b\gg a$ they are placed at $y \simeq \frac{a}{b}$ and $1-\frac{a}{b}$), corresponding to the closed and open subunit states respectively. $V_0$ is an energy scale associated with the barrier between both states, and $D$ is the energy difference between them.

{\bf -Interaction between gate sensor and membrane potential}. We consider that the voltage sensor of a subunit has a charge $Q$ that interacts with the membrane potential $\Delta V$. This can be modeled by the following energy potential, 
\begin{equation}
V_s(y,\Delta V)=Q(\Delta V-\Delta V_{ref})y,
\label{gateequ}
\end{equation}
where $\Delta V_{ref}$ is a reference potential.
Note that the parameters appearing in $V_g$ and $V_s$ could take different values for each gate if the channel had not a four-fold symmetry. That could be the case of modeling mutant K channels composed of subunits with different properties, such as those of the experiments in Ref.~\cite{chapman2005k}.

{\bf -Interaction energy between gates}. The coupling between the subunits of the same channel is modeled by an interacting term between pairs of subunits. We consider a square-like configuration (See Fig. \ref{tretramero}) in which each subunit interacts with its two nearest neighbors ($n.n.$), whereas the interaction with its opposed subunit is neglected. This is a ferromagnetic-like interaction favoring equal states.
The corresponding energy is given by
\begin{equation}
 V_c(\{y_j\})=\frac{\alpha}{2} \sum_{\{i,j\}_{n.n.}} \left( y_j - y_i\right)^2,
\end{equation}
where $\alpha$ represents the coupling energy scale, and the sum is over the four couples $\{i,j\}_{n.n.}$ of nearest neighbors interacting subunits.

\begin{figure}[ht]
\includegraphics[width=0.45\columnwidth,clip,angle=0]{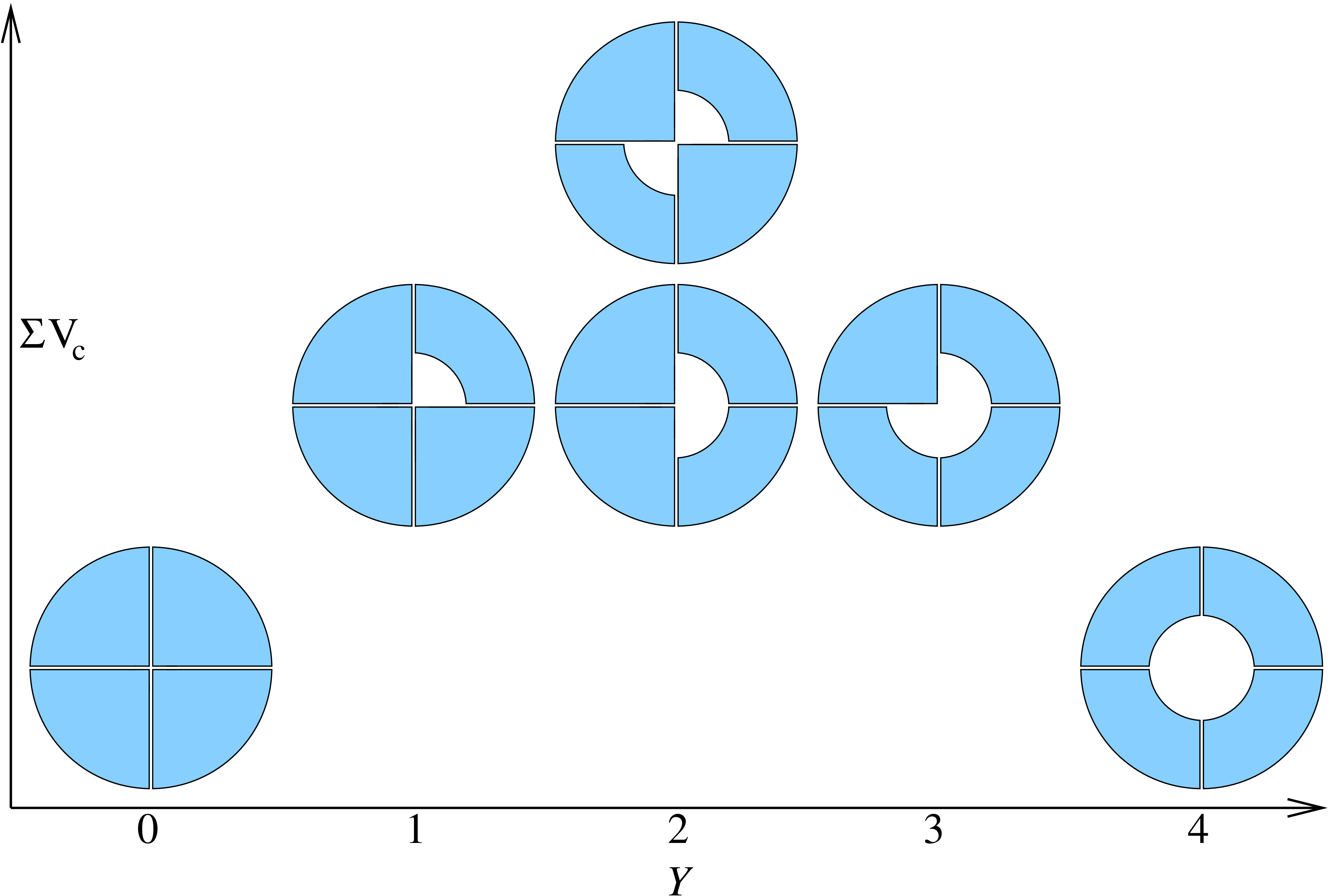}
\includegraphics[width=0.5\columnwidth,clip,angle=0]{Sumy-Vcoupling-corrected.eps}
\caption{Coupling energies $\sum V_c$ and order parameter $Y$ of the  different conformacional states of the channel. Left: schematic representation of each configuration. Right: samples of actual values during a long simulation of the sensorless pore. Note that the configuration with the highest energy appears very rarely along the simulation compared to the rest of the configurations. $\rho=1$ nm$^{-1}$, $\Delta V = 150$ mV. Sampled values are taken at intervals of 0.5 ms during a total run of 10 s.}
\label{Ecoupling}
\end{figure}

In Fig.~\ref{Ecoupling}-left we schematically represent all the possible configurations of the subunits, placed on the plot according to the value of this last coupling energy term $V_c$ and the value of the configurational order parameter $Y$. Two of these states, corresponding to the channel being completely open and completely closed, have a lower value of this term, and therefore they are energetically favored. 
It is also interesting to note that there are two different configurations with $Y\simeq 2$, that is two open subunits and hence with similar conductivities, but which have different energies. A similar plot, representing samples of actual values of $V_c$ and $Y$ during a simulation, is presented in Fig.~\ref{Ecoupling}-right. We can see in this plot that the more energetic state is visited much more rarely than the other three intermediate states, which have similar values of the coupling energy.

\begin{figure}[ht]
\includegraphics[width=0.8\columnwidth,clip,angle=0]{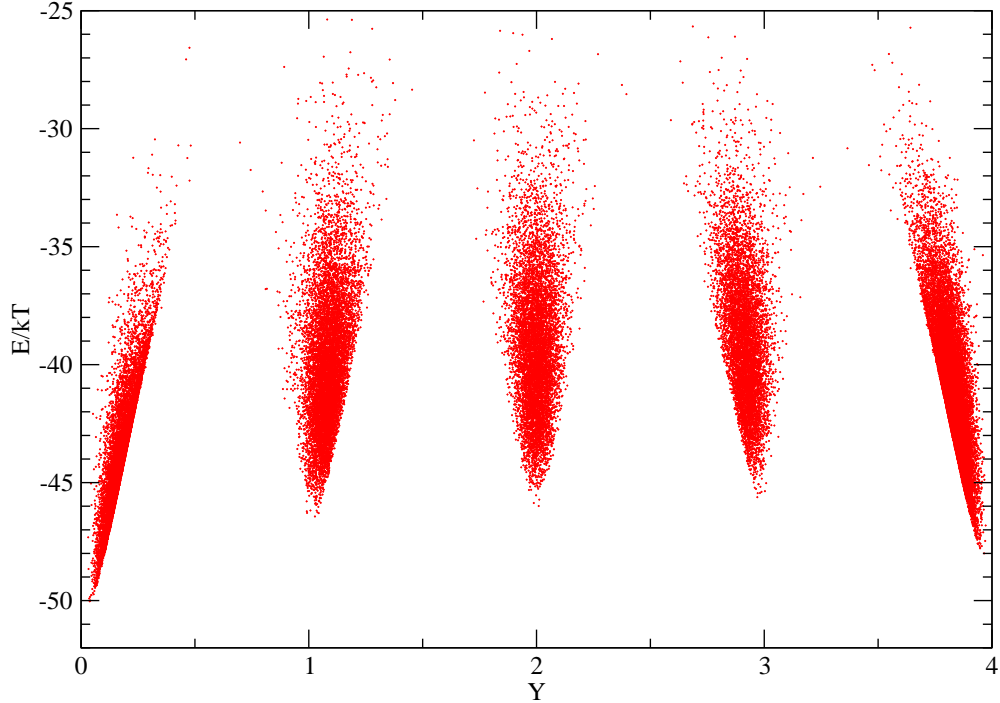}
\caption{Gate energy values $\sum Vg+\sum V_I+\sum V_c$, and order parameter $Y$, during a long simulation of the sensorless pore. $\rho=1$ nm$^{-1}$, $\Delta V = 150$ mV. Sampled values are taken at intervals of 0.5 ms during a total run of 10 s. 
}
\label{Ecoupling-alt}
\end{figure}

Taking together the energy terms corresponding to the gates, we have represented for illustrative purposes samples of $\sum Vg+\sum V_I+\sum V_c$ values versus the order parameter $Y$ in Fig.~\ref{Ecoupling-alt}. We observe large fluctuations in energy values, but still five groups of points are clearly distinguished, corresponding to the five main values of the order parameter $Y$. The sixth configurational state, that of the larger coupling energy, cannot be distinguished in this plot. Interestingly, despite the large fluctuations, the minimum energy value for each group is quite well defined. These minimum values are related to some parameters in the simulation. Indeed, the energy difference between all closed or all open states and the intermediate states is approximately 3 $k_BT$, that is the value $\alpha$ as it would be expected from the $V_c$ term. And over these values there is added a constant slope of roughly 0.5 $k_BT$ per $Y$ unit, which would correspond to the parameter $D$ in $V_g$.

\clearpage

\section{Intensities for subconductance states} 
\label{app-intensities}
Let us consider a one-dimensional system $(0,L)$ where Brownian particles move under a potential $V(x)$.
In the overdamped regime, the stochastic dynamical equation is
\begin{equation}
 {\dot x}= - \frac{V'(x)}{\gamma} +  \frac{\eta(t)}{\gamma},
\label{langevin}
\end{equation}
where  the  thermal noise $\eta(\tau)$ has  the correlation,
\begin{equation}
 \langle \eta(t)\eta(t') \rangle = 2 \gamma k_BT \delta(t-t').
\end{equation}

The corresponding Fokker--Planck equation for the density of particles  $\rho$ is,

\begin{equation}
 \frac{\partial \rho}{\partial t}=  \frac{\partial}{\partial x} \frac{1}{\gamma} \left[ V'(x) + k_BT\frac{\partial }{\partial x}  \right]\rho= -\frac{\partial}{\partial x } J
\label{FP3},
\end{equation}
where $J$ is the flux.

In the case of ions moving along a channel, we assume that the potential depends on the configurational state, and in particular on the number $n=0,1\dots 4$ of open subunits. We can write this potential as
\begin{equation}
V_n(x)=  -\frac{\Delta V}{L} x + U(x, n),
\end{equation}
where $\Delta V$ is the membrane potential and $U(x,n)$ depends on $n$. This term is given by (see Eq.~(\ref{VS}) and Fig.~\ref{barreras} in the main text)
\begin{equation}
 U(x,n) 
 =-k_BT e^\frac{-(x-\bar x)^2}{2\sigma^2}\ln[(1-\beta)(1-{\frac{1}{4}}F(n))].
\end{equation}
Here, the function $F(n)$ controls the reduction of the available section on the channel, depending on the number $n$ of open subunits:
\begin{equation}
 F(n)= n f_0 + (4-n)f_c,
\end{equation}
where $f_0$ is the value of $f(y)$ for an open state (see Eq.~\ref{eq:f} in Appendix \ref{app-potential}) and $f_c$ is the value for a close state. 

Given these conditions, we can find an expression for the expected flux. In the steady state, the flux obeys the equation 

\begin{equation}
 \frac{1}{\gamma} \left[ V_n(x) +
 k_BT \frac{d }{d x} \right] \rho = - J_n.
\label{J1}
\end{equation}
This is a linear non-homogeneous equation that can be easily solved. The formal solution is
\begin{equation}
 \rho_{st}(x)= \left( \rho_{st}(0) e^{-V_n(0)/k_BT} - \frac{J_n \gamma}{k_BT} \int_0^x e^{V_n(x')/k_BT} dx'\right) e^{-V_n(x)/k_BT}. 
\end{equation}

Imposing the boundary conditions at the edges $\rho_{st}(0)=\rho_{st}(L)=\rho$  we get the  flux,

\begin{equation}
J_n = \frac{k_BT \rho}{\gamma}\,\frac{e^{V_n(0)/k_BT}- e^{V_n(L)/k_BT}}{ \int_0^L e^{V_n(x')/k_BT} dx'},
\end{equation}
where the integral has to be obtained numerically. To perform this calculation we can make the change of variables  $z=x/L$, so the integration domain is now $(0,1)$. The flux is then,
\begin{equation}
J_n = \frac{k_BT \rho}{\gamma L}\,\frac{e^{V_n(0)/k_BT}- e^{V_n(L)/k_BT}}{ \int_0^1 e^{V_n(z)/k_BT} dz},
\end{equation}

Taking the values used in the simulation, the prefactor is
\begin{equation}
J_0=\frac{k_BT \rho}{\gamma L} =  12.5 \, \mu \text{s}^{-1},
\end{equation}
which is expressed as electrical intensity by using the ion charge $e=0.1602\times 10^{-7}$ pC as
\begin{equation}
 I_0=J_0 e = 2.0025 \,\text{pA}.
\end{equation}
The final explicit expression, including all information, to obtain the intensity is
\begin{equation}
 I_n(\Delta V) = I_0 \frac{e^{V_n(0)/k_BT}- e^{V_n(L)/k_BT}}{ \int_0^1 e^{V_n(z)/k_BT} dz}
 \label{theory-I-V}
\end{equation}
which is plotted in the main text (Fig.~\ref{int-y-y0.5a3}).
\end{appendix}
\end{document}